\documentclass{elsart}
\usepackage{graphicx}
\usepackage{amssymb}
\begin{document}
\begin{frontmatter}
\title{Persistence in an antiferromagnetic Ising system with conserved magnetisation} 
\author{Moumita Saharay and Parongama Sen }
\address{Department of Physics, University of Calcutta,
92 A.P. C. Road, Calcutta 700009, India.\\
e-mail: mouremi@yahoo.co.in; parongama@vsnl.net, paro@cubmb.ernet.in}
\begin{abstract}

We obtain the  persistence exponents for an antiferromagnetic Ising system
in which the magnetisation is kept constant. This system 
belongs to Model C (system with non-conserved order parameter
with a conserved density) and is expected to have  persistence exponents
different from that of Model A (system with no conservation) but independent
of the conserved density. Our numerical results for both local
persistence at zero temperature and global persistence at the
critical temperature  however indicate that
the  exponents are  dependent on the conserved magnetisation
in both two and three dimensions.
This nonuniversal feature is attributed to the presence of the conserved
field and is special to the persistence phenomena.

\end{abstract}

\end{frontmatter}

Persistence is a phenomenon \cite
{derrida,satyarev}
which occurs in  non-equilibrium dynamical systems
characterised by  the probability  $P(t)$ 
that a 
fluctuating field has not changed sign
upto the  time $t$. In systems exhibiting persistence,  $P(t)$  has a power 
law decay:
$P(t) \sim t^{-\theta}$.
Persistence occurs in many non-equilibrium dynamical processes like simple diffusion,
 reaction-diffusion systems, coarsening and  ordering kinetics  etc. Two kinds of persistence behaviour
 can be defined: {\em{local}}  and  {\em {global}},
related to the behaviour of the persistence probability of the
local and the global fields respectively.
In Ising models, local persistence is a much  studied problem  
where one studies the probability that the spin 
(the local fluctuating field) has not 
changed its sign
up to time $t$  when the system
is quenched to zero temperture from a high temperature.
For a quench to the 
critical 
temperature,  
global persistence is  observed 
 \cite{satya_global} when the probability 
that the global order parameter (e.g., magnetisation in a ferromagnetic system) has  not changed its  sign upto
time $t$ at the critical temperature, follows a power law decay.
Both the  exponents  $\theta_l$ associated with
the local persistence probability $P_l(t)$ and $\theta_g$ 
   associated with
the global  persistence probability $P_g(t)$
are new exponents not related to any previously known
exponent.

We have calculated numerically the local and global persistence exponents
in an antiferromagnetic Ising system in which the magnetisation is kept 
constant.  
This model belongs to the class of model C according to the classification 
\cite{Hohalp} of critical dynamical systems.
The dynamical critical  classes A, B, C etc. are distinguished by the long time 
dynamical exponent $z$ 
relating the divergences
in space and time.
Among the classes with  dissipative dynamics,    
 model A  is without any conservation while   the order parameter field in model B is conserved.
In model C, the order parameter field is not conserved, but it is  coupled to 
a non-ordering field which  is conserved. 

These  classes have characteristic values of the  dynamical exponents 
associated  with 
the different dynamical processes.
A short discussion on the persistence phenomena in these different systems
is necessary for understanding the results in the antiferromagnetic Ising system
with conserved magnetisation.
Local persistence phenomenon is quite well studied 
 in Model A. It is exactly solved for one dimension \cite{locmoda} 
 and extensively studied numerically for higher dimensions \cite{stauffer}.
The global exponent  for  model A
was obtained analytically    \cite{satya_global,oerding} and 
in subsequent numerical
simulations \cite{zheng3}.
In model B, the local order parameter shows persistence behaviour and the
persistence exponents using both local and global conservation
are  found to be non-universal \cite{ruten}.
The calculations were done for  phase ordering
dynamics and the exponents were found to vary linearly with the 
volume fraction of the minority phase which  is conserved.
Such non-universality does not occur in the long time dynamical
exponent $z$ in model B with local conservation.
Non-local conservation 
 gives rise to non-universality 
in some dynamical phenomena 
\cite{bray,satyacire,PS,zheng}. However, even in these cases,   the globally conserved case 
becomes identical 
to the non-conserved model with a unique exponent.  
Thus the non-universality of the exponents in model B is not related
to  the nature of  conservation but seems to be  a distinguishing feature of
persistence phenomena in a conserved system.
As the order parameter is conserved,  global persistence is not
defined in model B.

In model C one also has a conserved density,  but here 
 both local and global persistences  can  be considered.
Thus it is possible to  investigate here the
role of conservation 
in  both kinds of  persistences.
Estimate of the global persistence was made in \cite{oerding} up to
the second order in $\epsilon$, where $\epsilon$ is related to the 
dimensionality $d$ by $\epsilon = 4-d$.
Interestingly,  the global persistence
exponent
was found to be  universal (it does  not depend on the
non-ordering conserved density) unlike  the local persistence exponents in model B.
Our intention is to
verify the result of \cite{oerding} numerically and try to resolve the issue of
non-universality in conserved sytems for
both kinds of persistence behaviour in an appropriate model.

Conservation in models   B and C brings up the
question of non-universality. 
 On the other hand, since Model C
becomes identical to model A in certain limits, it is also
important to   address some general issues
concerning the dynamics in models A and C.

According to  analytical calculations, 
both  the long time and short time exponents in   models A and C  
are equal in two dimensions \cite{Hohalp,JSS,OJ} as the  specific heat exponent is zero. 
In three dimensions these are different. 
Numerical simulations \cite{SDS,sd} also agree with these results. 
The global persistence exponents in models A and C, however, are  different
even in two dimensions according to \cite{oerding}. 
If one assmues persistence to be a Gaussian Markov process, 
a scaling law relating
$\theta_g$ to other exponents can be derived \cite{satya_global}, 
which reads
\begin{equation}
\theta_g z=\lambda -d+1-\eta/2,
\end{equation}
where $\eta$ is a critical  exponent (related to the correlation function
$g(r)$ which behaves as $ r^{-(d-2+\eta)}$ for large spatial distance $r$ at the critical point) 
and  $\lambda$ is the exponent occurring in the scaling of the 
auto-correlation function,
$\langle A(t)A(0)\rangle  \sim t^{-\lambda}$,
where $A(t)$ is the field fluctuating in time.
Inserting the values of $\lambda$ and $z$ 
of  model A and model C, it can be shown that $\theta_g$
in both cases is given by
\begin{equation}
\theta_g=\frac{1}{2} -\frac{1}{12} \epsilon +O(\epsilon^2), 
\end{equation}
which implies they are equal  in all dimensions upto $O(\epsilon)$.
 This is however not true 
as  persistence is a non-Markovian phenomenon, and therefore,
global persistence exponents in models A and C
are different. 
We expect that local persistence exponents will also have different
values in these two models.

In our system the  order
parameter is the staggered magnetisation.  The  magnetisation is kept
constant but the order parameter is not conserved.
This system has earlier been considered as a candidate of model C
in context of the relaxation phenomena \cite{eisen}  and the  results 
obtained from the numerical  studies \cite{SDS,sd} in this model agree with the expected theoretical 
results. We have considered lattices of size  
$L$$\times$$(L+1)$ in two 
dimension and $L$$\times$$L$$\times$$(L+1)$ in three dimension as in \cite{sd}
 with helical boundary condition. The lattice size $L$ is kept 
odd as the system is antiferromagnetic.
 We have obtained the values of $\theta_l$ at zero temperature and $\theta_g$ at the critical temperature in 
both two  and three  
dimensions for fixed values of magnetisation $m_0$.

{\em Local persistence}: For local persistence, we have considered an initial state
where the order parameter ($m_s$) assumes zero value (i.e., the system is  prepared
as a random configuration). We consider zero temperature 
deterministic  Kawasaki dynamics which keeps the 
magnetisation  constant.
 It may be
noted that in  Glauber dynamics (e.g., in a 
ferromagnetic Ising case which belongs to the class of model A) the local spins 
are allowed to 
flip only once in a sequential sweep. In the Kawasaki dynamics on the other hand,
a spin may be flipped a number of times within a sweep, but we will consider a
change in the local spin  only if it  is different from its state in the previous time step 
after the   sweep has been completed.

This system becomes  equivalent to model A system when 
$m_0 = 0$ \cite{eisen}. In two dimenions, we first check that the value of $\theta_l$ in our model coincides with the
numerical value  $\sim 0.23$ of model A when the magnetisation 
$m_0$ is fixed at a zero value. As $m_0$ is made different from
zero, we find that the value of the exponent $\theta_l$ varies with $m_0$.
However, the behaviour with $m_0$ is not monotonic, 
$\theta_l$ first shows a dip for small $m_0$.    

\begin{figure}[t]
\begin{center}
\vskip -3cm
\includegraphics[width=15cm]{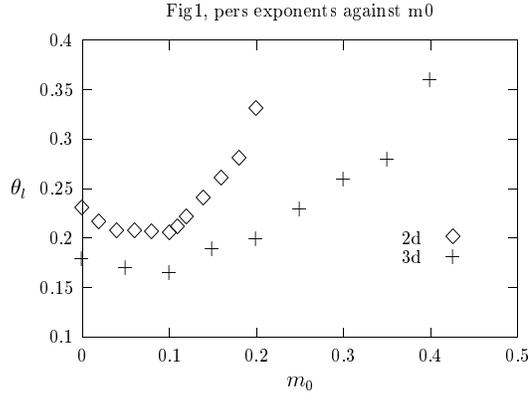}
\end{center}
\vskip -12.5cm
\caption{The variation of the local persistence exponent
with $m_0$ is shown in two and three dimensions.}
\label{fig1}
\end{figure}

In three dimensions, again the agreement of $\theta_l$ with the model A value of $\sim 0.18$ is
checked at $m_0 = 0$.
Here also we find that the persistence exponent is dependent
on $m_0$. As in two dimensions, it  initially shows a dip and then increases with $m_0$.
However, in the region  $0 < m_0 < 0.2$,
where it shows a dip, the distribution $P_l(t)$ shows deviation from
the power law behaviour  and therefore the estimates
of $\theta_l$ in this region may not be very accurate.
We have checked that the  results
are not due to finite size effect by studying systems of different sizes.
The behaviour of $\theta_l$ as a function of $m_0$ in two
and three dimensions is shown in Fig. 1 and the behaviour of $P_l(t)$ against $t$ for a few values of $m_0$ is shown in
For intermediate  values of $m_0$, the exponent   appears to vary linearly with $m_0$
in both two and three dimensions and
for higher values   of $m_0$, deviations from the power law start appearing in the
	probability distribution $P_l(t)$.

The above results cannot be compared to any  
available analytical study  but we note
 a few interesting features. First of all,
 the persistence exponent
 shows a variation with the conserved density while  other dynamical exponents
 in model C are universal \cite{Hohalp,OJ}.
 The non-universality is rather similar to the result of model B \cite{ruten}.
 Secondly, 
  the variation with $m_0$ is not smooth - the exponent first
  drops with $m_0$ and then increases.
  The deviation from the power law decay at higher $m_0$ could be
  attributed to the fact that the critical temperature 
  goes down to zero here.


\begin{figure}[t]
\begin{center}
\vskip -3cm
\includegraphics[width=12cm]{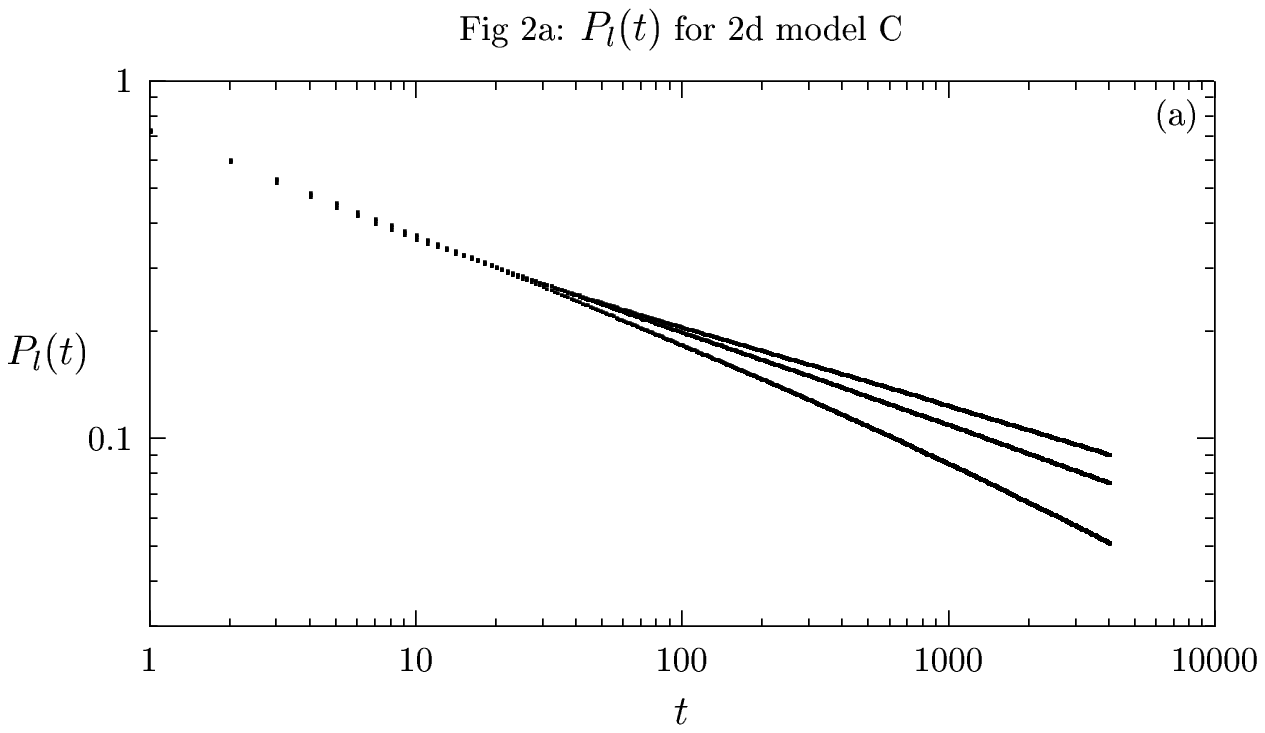}
\end{center}
\label{fig2a}
\end{figure}
\begin{figure}[t]
\vskip -13cm
\begin{center}
\includegraphics[width=12cm]{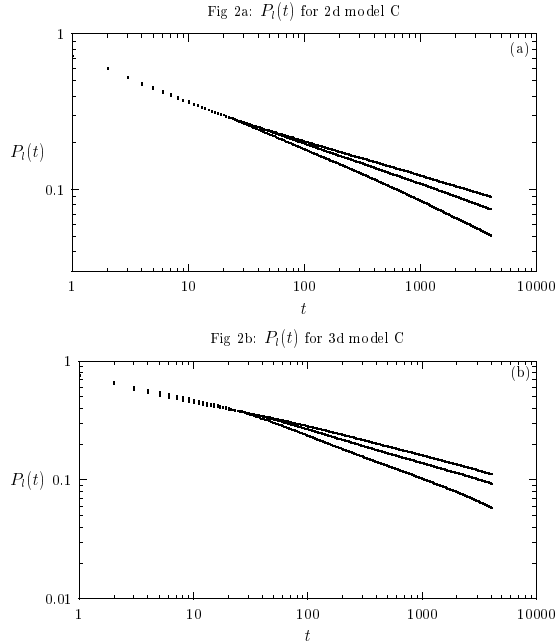}
\end{center}
\vskip -10cm
\caption{(a) The local persistence probability in two dimensions ($L=199$,
averaged over 100 initial configurations) for
$m_0$ = 0.12, 0.16 and 0.20 (from top to bottom) are shown.
(b) The local persistence probability in three dimensions  ($L=65$,
averaged over 100 initial configurations) for
$m_0$ = 0.30, 0.35 and 0.40 (from top to bottom) are shown.}
\label{fig2b}
\end{figure}

{\em Global persistence:}             
In our study of global persistence in this model, we considered two 
different kinds of initial conditions.
One is a sharply prepared (SP) state, where  the staggered magnetisation 
per site, i.e.,  $m_s(0)$ was kept fixed at small
 positive  values (0.0001,
0.0002, 0.0005 etc.) initially. Ideally, $\theta_g$ for each of these 
values are to be  
calculated and the result extrapolated to $m_s(0) = 0$.
Another initial condition corresponds to a  
randomly 
prepared (RP) state, where the staggered
magnetisation assumes a small value initially which can be either positive or negative.
For both states the magnetisation is kept constant at a fixed value $m_0$. 
The dynamics is again Kawasaki but it is stochastic.
          
          Since the system is quenched to the critical temperature $T_c(m_0)$,
we have to know the values of $T_c(m_0)$. We have used the values of $T_c(m_0)$ given in 
\cite{sd}.

\begin{figure}[t]
\begin{center}
\vskip -3cm
\includegraphics[width=12cm]{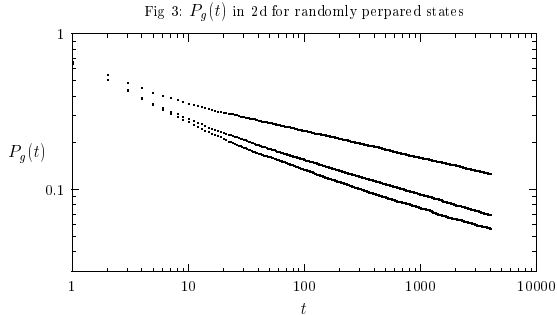}
\end{center}
\vskip -10cm
\caption{The global persistence probability in two dimensions for random initial
conditions ($L$=199)
for $m_0$
	 = 0.2, 0.1 and 0 (from top to bottom) are shown.
	 Typically 3$\times 10^4$ initial configurations are generated.}
	 \label{fig3}
	 \end{figure}

 We have obtained the values of
$\theta_g$ for several values of non zero $m_0$ and also for $m_0$=0 
for comparison with model A. 
Note that for the global persistence one needs the fraction of configurations for which the order parameter does not change sign and to
get a good estimate of $\theta_g$ one needs to generate 
a large number of configurations. Thus to restrict the 
total computation, we have evaluated $\theta_g$ for a few values of 
$m_0$ only unlike the local case. Moreover, in the global case, the system is  
critical and estimating the critical temperature again 
involves a lot of computations.

\begin{small}
\begin{center}

{\bf  Table 1 } \\
Numerical estimates of global persistence
exponent $\theta_g$ for randomly prepared (RP)
and sharply prepared (SP) states in
an antiferromagnetic Ising system
with conserved magnetisation.
In each  case the error
is $\pm 0.01$\\
\medskip

\end {center}
\begin{center}

\begin{tabular} {|c|c|c|c|} \hline
Dimension &$m_0$&Estimate         &Estimate \\
($d$)  &      &of $\theta_g$ for&of $\theta_g$ for \\
       &      &SP               & RP \\ \hline
  2    & 0.0  & $\sim  0.23$    & $\sim 0.24$ \\
      &0.1   & $\sim 0.23$     & $\sim 0.23$ \\
      &0.2   & $\sim 0.18$     & $\sim 0.17$ \\ \hline
  3    & 0.0  & $\sim 0.40$     & $\sim 0.40$ \\
       & 0.08 & $\sim 0.40$     & $\sim 0.39$ \\
       & 0.2  & $\sim 0.35$     & $\sim 0.33$ \\
       & 0.4  & $\sim 0.31$     & $\sim 0.30$ \\ \hline
\end{tabular}

\end{center}

\end{small}

The numerical
values of $\theta_g$ for model A \cite{stauffer} are recovered when $m_0=0$
with both RP and SP   initial conditions. 
In both two and 
three dimensions,  the values of $\theta_g$  appear to depend 
on $m_0$ (see Table I).
	In the SP  
initial condition, estimation of $\theta_g$ becomes 
difficult as the fluctuations are considerable even with $3\times 10^4$
configurations.
Hence instead of the extrapolation procedure, we 
give in  Table I the
value of $\theta_g$ for $m_s(0) = 0.0005$  and $m_s(0) = 0.0002$ in
two and three dimensions respectively  for which the fluctuations are minimum.
For low values of $m_0$, $\theta_g$ is very close (if not equal) to
the model
A values  with both RP and SP states. For higher values of $m_0$,
it is different from that of model A and seems to vary with $m_0$.
Typical variations of $P_g(t)$ in two and three dimensions are  shown in Fig. 3 and 4.
In agreement with \cite{oerding}, we indeed find that the model A and model C 
global persistence exponents are unequal
even in two dimensions (if we use a large value of $m_0$),
although  in contrast to \cite{oerding} we find non-univerality.
This discrepancy might arise from the fact that in \cite{oerding} 
a continuum model was considered while we have considered here a discrete system.
Such discrepancies have been observed for the zero-temperature coarsening dynamics of the continuum 
model B and the corresponding discrete model \cite{cueille}.

\begin{figure}[t]
\begin{center}
\vskip -3cm
\includegraphics[width=12cm]{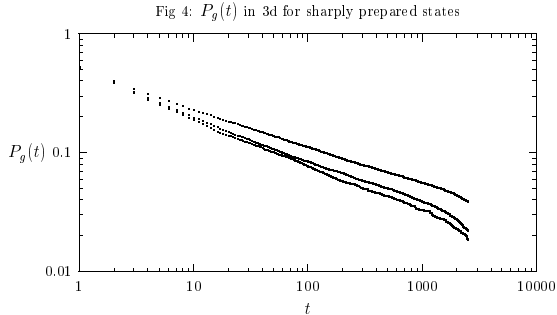}
\end{center}
\vskip -10cm
\caption{Global persistence probability for sharply prepared initial condition
in three dimensions for
$m_0$
= 0.4, 0.2 and 0 (from top to bottom) with $L$=41 are plotted.
Typically 3$\times 10^4$ initial configurations are generated.}
\label{fig4}
\end{figure}

The analytical results of model B and the present numerical results
in a system belonging to model C suggest that in persistence phenomena 
conservation
plays a key role which may be  responsible for the non-universality
of the exponents and it is not important whether  conservation
is considered in the order parameter or in a coupled non-ordering field.
However, the analytical study of \cite{oerding} and a 
recent study \cite{arko} of both the local and global persistence 
in an absorbing phase transition in a conserved model does not show 
non-universality.
Therefore for a better understanding of our results it is necessary to get
an analytical estimate of the 
local persistence behaviour in model C. Also, the global persistence
may be studied in greater detail both analytically and numerically. 
The non-monotonic behaviour in the local persistence exponent
for small $m_0$ is also a novel observation.
In the case of global persistence, owing to
the large number of configurations required to get a reliable result,
 we had to restrict the computations to a few values of $m_0$
and it was not possible to study in detail the behaviour for small $m_0$ 
where the local persistence exponent shows the anomalous behaviour.

It is also interesting to compare the local and global persistence
in models A and C.
In model A, in three dimensions, $\theta _g > \theta_l$. 
From the present study, we find that while $\theta_l$ increases 
 for $m_0$ above a certain value, $\theta_g$ decreases 
with $m_0$. Hence the relation $\theta_g > \theta_l$ breaks down 
above a certain $m_0$ in model C. The exact significance of this result
is yet to be understood.

Lastly, all the above results for local and glocal persistence 
in the antiferromagnetic Ising system with conserved magnetisation 
were obtained with local conservation. It is expected that non-local
conservation will lead to different results.

Acknowledgements: We are greatful to S. Dasgupta for valuable discussions. 
PS acknowledges DST grant no SP/S2/M-11/99.
The computations were done on an Origin200 in CUCC.


\begin{thebibliography}{99}
\bibitem{derrida}
A. J. Bray, B. Derrida and C. Godreche, J. Phys. {\bf A 27}, L357 (1994);
\bibitem{satyarev} For a review, see S. N. Majumdar, Curr. Sci. India {\bf 77} 370 (1999).
\bibitem{satya_global} S.N.~Majumdar, A. J. Bray, S. J. Corwell and C. Sire,  
 Phys. Rev. Lett. {\bf 77} 3704 (1996).
\bibitem{Hohalp} P.C.~Hohenberg and B.I.~Halperin, Rev. Mod. Phys. {\bf 49}
435 (1977).
\bibitem{locmoda}B. Derrida, V. Hakim and V. Pasquier, Phys. Rev. Lett. 
{\bf 75} 751 (1995).
\bibitem{stauffer}A. J. Bray, B. Derrida and C. Godreche, Eur. Phys. Lett. 
{\bf 27} 175 (1994); 
D. Stauffer, J. Phys. A {\bf 27} 5029 (1994);
S. N. Majumdar and C. Sire, Phys. Rev. Lett. {\bf 77} 1420 (1996).
\bibitem{oerding} K. Oerding, S. J. Cornell and A. J. Bray, 
Phys. Rev. E {\bf 56}
 R25 (1997).
\bibitem{zheng3} 
 L. Schuelke, B. Zheng, Phys. Lett. A {\bf 233} 93 (1997). 
\bibitem{ruten}B. P. Lee and A. D. Rutenberg, Phys. Rev. Lett. {\bf 79}
4842 (1997).
\bibitem{bray} A.J.~Bray, Phys. Rev. Lett. {\bf 66} C2048 (1991).
\bibitem{satyacire} C.~Sire and S.N.~Majumdar, Phys. Rev. E {\bf 52} 244 (1995).
\bibitem{PS} P.~Sen, J. Phys. A {\bf 32} 1623 (1999).
\bibitem{zheng} B.~Zheng and H.J.~Luo,  Phys. Rev. E {\bf 63} 066130 (2001).
\bibitem{JSS} H. K. Janssen, B. Schaub and B. Schmittmann, Z. Phys. {\bf B73}
539 (1989).
\bibitem{OJ} 
 K.~Oerding and H.K.~Janssen, J. Phys. A {\bf 26} 3369 (1993).
\bibitem{SDS} P.~Sen, S.~Dasgupta and D.~Stauffer, Eur. Phys. J. B {\bf 1}
107 (1998).
\bibitem{sd}P.Sen and S.Dasgupta, J. Phys. A {\bf 35} 2755 (2002).
\bibitem{eisen} E. Eisenriegler and B. Schaub, Z. Phys. B {\bf 39} 65 (1980).
\bibitem{cueille} S. Cueille and C. Sire, J. Phys. A {\bf 30} L791 (1997).
\bibitem{arko} S.  Lubeck and A. Misra, Eur. Phys. J. B,  to be published, 
(cond-mat/0201411). 
\end{thebibliography}
\end{document}